\documentclass[12pt,dvips]{article}

\usepackage[english]{babel}
\usepackage{amsfonts}
\usepackage{amssymb}
\usepackage{amsmath}
\usepackage{graphicx}
\usepackage{graphics,wrapfig}

\begin{document}

\title{
Pion and Quark Annihilation Mechanisms of Dilepton Production in
Relativistic Heavy Ion Collisions }

\author{
D.~Anchishkin$^{a,}$\footnote{E-mail:
anch@bitp.kiev.ua} ,
R.~Naryshkin$^{b,}$\footnote{E-mail: Naryshkin@univ.kiev.ua}\\
{\small \it $^a$Bogolyubov Institute for Theoretical Physics,
           Kyiv 03143, Ukraine}\\
{\small \it $^b$
Kyiv National University, Physics Department,
Kyiv 03022, Ukraine }
}

{\small \date{}}

\maketitle


\begin{abstract}
\noindent
We revise the $\pi^+\pi^-$ and $q\bar{q}$ annihilation mechanisms of
dilepton production during relativistic heavy-ion collisions. We focus
on the modifications caused by the specific features of intramedium
pion states rather than by medium modification of the $\rho $-meson
spectral density. The main ingredient emerging in our approach is a
form-factor of the multi-pion (multi-quark) system. Replacing the
usual delta-function the form-factor plays the role of distribution
which, in some sense, "connects" the 4-momenta of the annihilating
and outgoing particles. The difference between the c.m.s. velocities
attributed to annihilating and outgoing particles is a particular
consequence of this replacement and results in the appearance of a
new factor in the formula for the dilepton production rate. We
obtained that the form-factor of the multi-pion (multi-quark) system
causes broadening of the rate which is most pronounced for small
invariant masses, in particular, we obtain a growth of the rate for
the invariant mass below two masses of the annihilating particles.

\end{abstract}


\noindent
Since leptons do not
interact practically with the highly excited nuclear matter, they
leave the reaction zone
after their creation in a relativistic nucleus-nucleus collision
without further rescattering.
That is why, the dileptons ($e^+e^-$ and $\mu^+\mu^-$ pairs) observed
in high-energy heavy ion collisions secure an excellent opportunity
for obtaining information on the initial state and the evolution of the
system created in a collision.
The enhancement with an
invariant mass of $200\div 800$ MeV observed by the CERES
collaboration \cite{1,2} in the production of dileptons has received
recently a considerable attention and has been studied in the
framework of various theoretical models (for the review, see Ref.
\cite{3}).
It was found that a large part of the observed enhancement is due to
the medium effects (see Refs. \cite{4,5} and references therein).
Meanwhile, pion annihilation is the main source of dileptons which
come from the hadron matter \cite{6,7}.
That is why, the proper analysis of the dilepton
spectra obtained experimentally gives important observables which
probe the pion dynamics in the dense nuclear matter that exists at the
early stage of the collision.
The purpose of the present letter is to look once more on the
$\pi^+\pi^-$ annihilation mechanism of dilepton production from the
hadron plasma by accounting the medium-induced modifications of the
dilepton spectrum.
In order to do this, we concentrate on the modifications which are
due rather to intramedium pion states,
than on the discussion of a modification of the $\rho$-meson
spectral density.
In accordance with our suggestions, the main features of a
pion wave function follow from the fact that pions live a finite time
in the system where they can take part in the annihilation
reaction.
As a consequence, the off-shell effects give an appreciable
contribution to the features of the annihilation process
specifically in the region of low invariant masses.
Moreover, if pions are the entities of a local subsystem, then the
spatial structure of the pion states is far from a plane-wave one and
this also gives the essential contribution to the features of the
dilepton spectrum.

To carry out the outlined program, we assume that the pion liquid
formed after the equilibration exists in a finite
volume, and the confinement of pions to this volume is a direct
consequence of the presence of the dense hadron environment which
prevents the escape of pions during some mean lifetime $\tau$.
The same can be assumed concerning a hot system of quarks which are
confined to a quark-gluon droplet.
We sketch the proper geometry in Fig.~1.
The small circle of radius $R$ represents a subsystem of pions
which is in a local thermal equilibrium and moves with the
collective velocity $\mathbf{v}$.
If the fireball size is of order of
$R_0\propto 6\div 10$ fm, then the mean size of the small
subsystem is of order of $R\propto 1\div 5$ fm.
So, we assume the system of pions (quarks) produced in high-energy
heavy-ion collisions is effectively bounded in a finite volume.
The pion (quark) wave functions $\varphi_\lambda(x)$, where
$\lambda$ is a quantum number, satisfy the proper boundary conditions
and belong to the complete set of functions.
For instance, the stationary wave functions may be taken as the
solutions of the Klein-Gordon equation
$\left( \mathbf{\nabla} ^2 +k^2 \right)\varphi_\lambda({\bf x})= 0$,
where $k^2=E^2-m^2$, which satisfy the Dirichlet boundary condition
on the surface $S$: $\varphi_\lambda({\bf x})|_S=0$.
For the box boundary, we get
$\varphi_\mathbf{k}({\bf x})
=
   \sqrt{8/V}
   \prod_{i=1}^3  \theta(L_i-x_i) \theta(x_i) \sin{(k_i x_i)} $,
where $V=L_1L_2L_3$ is the box volume,
$\lambda\equiv {\bf k}=(k_1,k_2,k_3)$, and components of the
quasi-momentum
run through the discrete set $k_i=\pi n_i/L_i $ with
$n_i=1, 2, 3, \ldots$\, .
For the spherical geometry, the normalized solutions are written as
$\varphi_{klm}({\bf r})
=
   \theta(R-r)  \left( 2/r \right)^{1/2}
  J_{l+1/2} (kr)
  Y_{lm}(\vartheta,\phi)/
  RJ_{l+3/2} (kR)$,
where $\lambda=(k,l,m)$.
Next, the field operators $\hat{\varphi}(x)$ corresponding to the
pion field should be expanded in terms of these eigenfunctions, i.e.
\begin{equation}
\hat{\varphi}(x)=
\int\frac{d^3k}{(2\pi)^32\omega_\mathbf{k}}
\left[
a(\mathbf{k})\varphi_\mathbf{k}(x)+b^+(\mathbf{k})\varphi^*_\mathbf{k}(x)
\right]
\, ,
\label{1}
\end{equation}
where $a(\mathbf{k})$ and $b(\mathbf{k})$ are the annihilation
operators of positive and negative pions, respectively.
%
%
\begin{wrapfigure}{l}{0.3\textwidth}
{\includegraphics
     [width=0.3\textwidth,height=0.21\textheight]{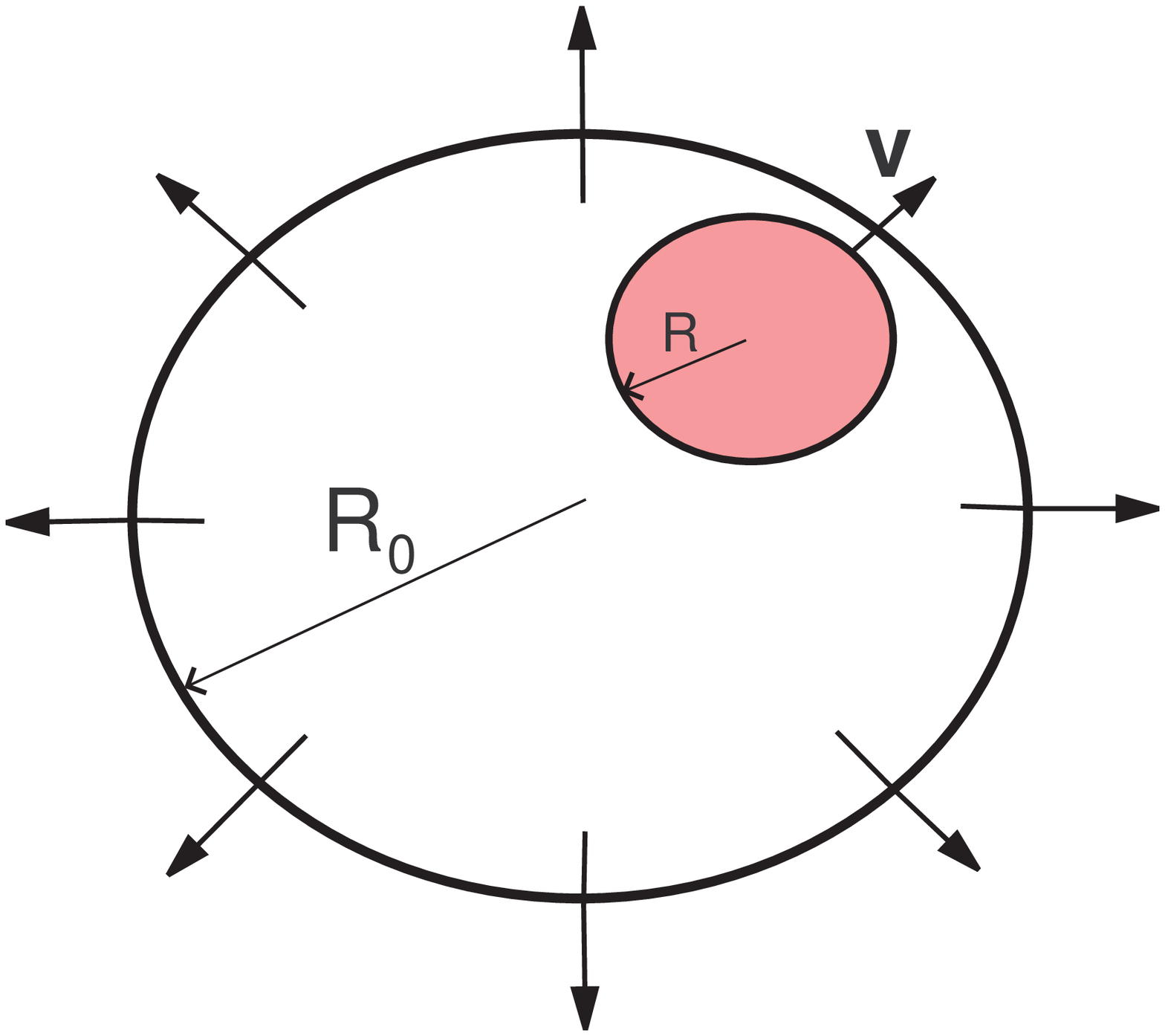}}
\noindent
{\footnotesize Fig.~1.
Sketch of an expanding fireball.
The small circle of radius $R$ represents the subsystem of pions
which is in a local thermodynamic equilibrium and moves with the
collective velocity ${\bf v}$.
}
\end{wrapfigure}
%
On the other hand, the states corresponding to
confined particles can be written in a common way as
$\varphi_\mathbf{k}({\bf x})
=
  \sqrt{ \rho({\bf x})/V } \, \Phi_\mathbf{k}({\bf x})$,
where
$\rho^{1/2}({\bf x})=
\prod_{i=1}^3 \left[ \theta(L_i-x_i)\theta(x_i)\right]$
for a box and $\rho^{1/2}({\bf x})=\theta(R-r)$ for a sphere,
respectively.
The function $\rho({\bf x})$ presents the information about the
geometry of a reaction region or cuts out the volume where the pions
(quarks) can annihilate.
Hence, for the evaluation
of $S$-matrix elements
wave functions $\varphi_\mathbf{k} (x)$ should be taken as
the pion $in$-states once annihilating pions belong to finite system.
The amplitude of the pion-pion annihilation to a lepton pair in the
first non-vanishing approximation is calculated via the chain
$\pi^+\pi^-\rightarrow \rho\rightarrow\gamma^*\rightarrow
\bar{l}l$, where
the $\rho$-meson appears as an intermediate state in accordance with
the vector meson dominance.
The matrix element of the reaction is
$\langle \mathrm{out} | S | \mathrm{in} \rangle
 =  - \int d^4x_1\, d^4x_2 \,
   \langle {\bf p}_+,{\bf p}_- \left|
   \,  T \left[  {\it H}_I^\pi(x_1) \, {\it H}_I^l(x_2)  \right] \,
   \right| {\bf k}_1,{\bf k}_2 \rangle   $,
where
${\it H}_I^\pi(x)=-e\, j_\mu^\pi(x) \, A^\mu(x)$
and
${\it H}_I^l(x)=-e\, j_\mu^l(x) \, A^\mu(x)$.
It is remarkable that the pion density $\rho({\bf x})$ appears as a
factor of the pion current.
Indeed,
\begin{equation}
j^\pi_\mu(x)
=
  -i \hat{\varphi} (x)
  \stackrel{\leftrightarrow}{\partial_\mu} \hat{\varphi}^+(x)
=
  \frac{\rho({\bf x})}{V}
  \left[ -i \hat{\Phi} (x) \stackrel{\leftrightarrow}{\partial_\mu}
     \hat{\Phi}^+(x) \right]
\, ,
\label{1a}
\end{equation}
where the field operator
$\hat{\Phi} (x)$ is defined in the same way as that in (\ref{1})
with the functions $\varphi_\mathbf{k}({\bf x})$ replaced by
$\Phi_\mathbf{k}({\bf x})$.
Because of this factorization, after the integration over the
vertex $x$ the density $\rho({\bf x})$ automatically cuts out the
volume, where the $\pi^+ \pi^-$ annihilation reaction is running.
At the same time, this means that the density $\rho({\bf x})$
determines the volume of quantum coherence, i.e. just the particles from
this spatial domain are capable to annihilate one with another and
make contribution to the amplitude of the reaction.
To obtain the overall rate, it is necessary then to sum up the rates
from every coherent domain of the fireball.

For the sake of simplicity, we assume that the pion states
can be approximately represented as
$\varphi_\mathbf{k}(x)=\sqrt{\rho(x)/V}e^{-i k\cdot x}$.
Here, $\rho(x)$ is the 4-density of pions in the volume $V$ (marked
by radius $R$ in Fig.~1), where pions are in a local thermodynamic
equilibrium.
In essence, this approximation considers just one mode of the wave
function $\Phi_\mathbf{k}({\bf x})$.
We expect that this ansatz reflects the qualitative features of the
pion states in a real hadron plasma, and, below, the main
consequences of the contraction of particle states in the
dense environment are considered.

A simple calculation immediately shows that the $S$-matrix element
is proportional to the Fourier-transformed pion density $\rho(x)$, i.e.
$\langle \mathrm{out} | S | \mathrm{in} \rangle \propto
 \rho(k_1+k_2-p_+-p_-)$,
where $k_1$ and $k_2$ are the
$4$-quasi-momenta of the initial pion states and $p_+$ and $p_-$ are
the $4$-momenta of the outgoing leptons.
This means that the form-factor of the pion source
$\rho(k)$ stands here in place of the delta function which appears in
the standard calculations, i.e.
$(2\pi)^4\delta^4(K-P)\rightarrow \rho(K-P)$,
where $K=k_1+k_2$ and $P=p_++p_-$  are the total (quasi-) momenta
of pion and lepton pairs, respectively.
An immediate consequence of this is a breaking down of the
energy-momentum conservation in the $s$-channel of the reaction, which
means that the total momentum $K$ of the pion pair is no longer equal
exactly to the total momentum $P$ of the lepton pair.
The physical
interpretation of this fact is rather obvious: the effect
of the hadron environment on the pion subsystem which prevents the
escape of pions from the fireball can be regarded during the time
span  $\tau$ as the influence of an external nonstationary field.
The latter, as known, breaks down the energy-momentum conservation.
From now, the squared form-factor $|\rho(K-P)|^2$ of the pion system
plays the role of a distribution which in some sense "connects" in
{\it s}-channel the annihilating and outgoing particles instead of
$\delta$-function.
Indeed, the number $N^{(\rho)}$  of produced lepton pairs from a
finite pion system related to an element of the dilepton momentum
space, reads
\begin{equation}
\left< \frac{dN^{(\rho)}}{d^4P}\right>=\int d^4K
|\rho(K-P)|^2 \left< \frac{dN}{d^4Kd^4P}\right>
\, ,
\label{2}
\end{equation}
where
\begin{multline}
\left< \frac{dN}{d^4Kd^4P}\right>
=
\int \frac{d^3k_1}{(2\pi)^3 2E_1} \frac{d^3k_2}{(2\pi)^3 2E_2} \,
     \delta^4(k_1+k_2-K)\, f_\mathrm{th}(E_1)\, f_\mathrm{th}(E_2)
\\
\times
\int \frac{d^3p_+}{(2\pi)^3 2E_+} \frac{d^3p_-}{(2\pi)^3 2E_-} \,
     \delta^4(p_++p_--P)|A_0(k_1,k_2;p_+,p_-)|^2
\, .
\label{3}
\end{multline}
Here, $E_i=\sqrt{m^2+\mathbf{k}_i^2}$, $i=1,2$ for pions and $i=+,-$
for leptons, respectively.
By the broken brackets, we denote the thermal
averaging over the pion quasi-momentum space with the
help of the thermal distribution function $f_\mathrm{th}(E)$.
To obtain eq.(\ref{2}), we represent the amplitude
$A_\rho=\langle \mathrm{out} | S^{(2)} | \mathrm{in} \rangle$
of the reaction under consideration as
\begin{equation}
A_\rho(k_1,k_2;p_+,p_-)
=
\rho(k_1+k_2-p_+-p_-) A_0(k_1,k_2;p_+,p_-)
\, .
\label{4}
\end{equation}
We note that not only the form-factor $\rho(K-P)$ contains
information about the pion system.
The amplitude $A_0$ carries new important features as well, which are
related to the violation of the energy-momentum conservation in the
$s$-channel.
Indeed, the pion-pion c.m.s. moves with the velocity
$\mathbf{v}_{K}=\mathbf{K}/K_0$, whereas the lepton-pair c.m.s.
moves with the velocity  $\mathbf{v}_{P}=\mathbf{P}/P_0$.
Hence, these two center-of-mass systems are "disconnected"{} now,
that is why any quantity should be Lorentz-transformed
when transferred from one c.m.s. to another.
A particular consequence of this is the appearance of the correction
factor
$\left[1+\frac{1}{3}\left(\frac{(P\cdot K)^2}{P^2K^2}-1\right)\right]$
in the formula for the dilepton production rate (for details see \cite{8}).

Concerning the physical meaning of eq.~(\ref{2}), we note that
one can regard it as the averaging of the random quantity
$\left<\frac{dN}{d^4Kd^4P}\right>$  with the help of the distribution
function $|\rho(K-P)|^2$ centered around the mean value $P$.
In this sense, the hadron medium holding pions in a local spatial
region for some time,
which is expressed as the local pion distribution $\rho(x)$,
plays the role of an environment randomizing the pion source.
This randomization is a purely quantum one in contrast to the thermal
randomization of the multi-pion system which is already included
to the quantity $\left< \frac{dN}{d^4Kd^4P}\right>$.

In order to transform the distribution of the number of created
lepton pairs over the dilepton momentum space to the
distribution over invariant masses, one has to perform
additional integration using $\left< dN^{(\rho)}/d^4P \right>$
from (\ref{2}), i.e.
$\left< \frac{dN^{(\rho)}}{dM^2} \right> = \int\frac{d^3P}{2P_0}
\left< \frac{dN^{(\rho)}}{d^4P} \right>$, where
$P_0=\sqrt{M^2+\mathbf{P}^2}$.
Taking this together we obtain
\begin{multline}
\label{5}
\left<\frac{dN^{(\rho)}}{dM^2}\right>
=
 \frac{\alpha^2}{3(2\pi)^8} \left(1-\frac{4m^2_e}{M^2}\right)^{1/2}
 \left(1+\frac{2m^2_e}{M^2}\right) |F_\pi(M^2)|^2
 \int\frac{d^3P}{2P_0}\int d^4K \frac{K^2}{M^2} \,
     \left|\rho(K-P)\right|^2 e^{-\beta K_0}
\\
 \times\left(1-\frac{4m^2_\pi}{K^2}\right)^{3/2}\left[1+\frac13
 \left(\frac{(P\cdot K)^2}{M^2K^2}-1\right)\right]\theta(K_0)
 \theta(K^2-4m^2_\pi)
\, ,
\end{multline}
where  $F_\pi(M^2)$ is the $\rho$-meson form-factor, and we take the
Boltzmann distribution $f_\mathrm{th}(E)=\exp(-\beta E)$ with
$\beta=1/T$ as inverse temperature.
By the presence of the
$\theta$-functions (last two factors on the r.h.s. of (\ref{5})),
we would like to stress that the invariant mass of a pion pair
$M_\pi=\sqrt{K^2}$ is not less than two pion masses.
On the other hand, possible finite values of the distribution
$\left<\frac{dN^{(\rho)}}{dM^2}\right>$ below the two-pion mass
threshold can occur just due to the presence of the pion system
form-factor $\rho(K-P)$.
%
\begin{wrapfigure}{l}{0.32\textwidth}
{\includegraphics[width=0.3\textwidth,height=0.22\textheight,
                  angle=-90]{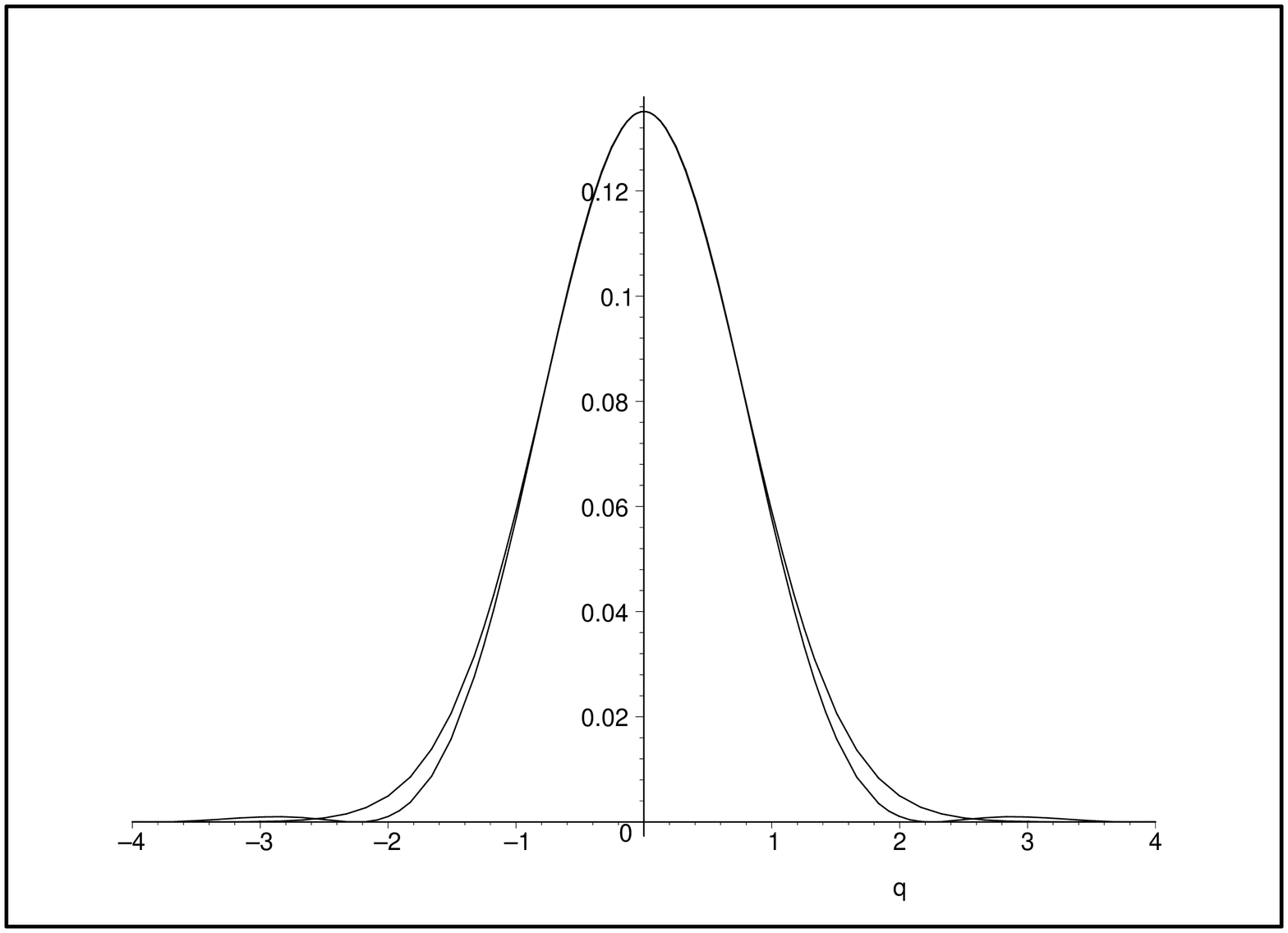}}

\vspace{30pt}
\noindent
{\footnotesize Fig.~2.
Comparison of the Gaussian and $\theta-$function form-factors
$|\rho(\mathbf{q})|^2/V$.}
\end{wrapfigure}
The factor in the square brackets on the
r.h.s. of (\ref{5}) is a correction which is due to the Lorentz
transformation of the quantity $(\mathbf{k}_1-\mathbf{k}_2)^2$
from the dilepton c.m.s. to the pion-pion c.m.s.
This factor gives a remarkable contribution to the dilepton spectrum
for invariant masses below the two-pion mass value.
Its influence is especially pronounced for $e^+e^-$ production
(see \cite{8}).

To clarify the effects under investigation as much as possible, for
particular evaluations we take as a model of pion system the Gaussian
distribution of the particles in space and the Gaussian decay
of the system of pions:
\begin{equation}
\rho(x)
=
\exp{\left(-\frac{t^2}{2\tau^2}-\frac{\mathbf{r}^2}{2R^2}\right)}
.
\label{6}
\end{equation}
Meanwhile, it can be another choice of the model function for the
pion source.
Indeed, one can choose for instance a geometry with sharp boundaries
which are determined by the $\theta$-functions.
To elucidate this point we compare two formfactors (normalized to the
unit volume) which correspond to the Gaussian distribution
$\rho_g(\mathbf{r})=e^{-\frac{\mathbf{r}^2}{2R^2}}$ and to the
$\theta$-function distribution
$\rho_\theta(\mathbf{r})=\theta(R-|\mathbf{r}|)$ (see Fig.~2).
Only a slight difference between these formfactors is seen and,
therefore, the choice of pion source distribution doesn't affect much
the dilepton production rate.
In what follows, we dwell on the Gaussian distribution.

To be closer to the quantities measured in experiment, we evaluate
the rate related to the proper rapidity window
$$\frac{dR}{dM\, dy}=2\pi M\frac{1}{\triangle y}
\int_{y_\mathrm{min}}^{y_\mathrm{max}}\, dy
\int_{P_{\perp}\mathrm{min}}^\infty dP_\perp\, P_\perp
               \frac{dN}{d^4x\, d^4P}\, ,$$
where $\tanh{y}=P^3/P^0,\quad P_\perp^2=(P^1)^2+(P^2)^2$.
The results of evaluation of the production rates
$dR^{(\rho)}_{e^+e^-}/dMdy$ and $dR^{(\rho)}_{\mu^+\mu^-}/dMdy$ for
electron-positron and muon-muon
pairs, respectively, in pion-pion annihilation are depicted in Fig.~3.
Different curves correspond to the different "spatial sizes" $R$ and
different "lifetimes" $\tau$  of a hot pion system at the
temperature $T=180$ MeV.
Notice that the production rate in a finite small pion system
differs from the rate in a infinite pion gas (solid curve) where
pion $in$-states can be taken as plane waves.
The deviation bigger when the parameters $R$ and $\tau$ are smaller.
Of course, this is a reflection of the uncertainty principle which is
realized by the presence of the distribution
$|\rho(K-P)|^2$ as the integrand factor in (\ref{2}).
Basically, the presence of the form-factor of the multi-pion
system will result in a broadening of the rate for small
invariant masses $M\leqslant 800$~MeV/c$^2$ which is wider at the
smaller parameters $R$ and $\tau$.
This seems natural because the quantum fluctuations of the momentum
are more pronounced in smaller systems.
We emphasize as well that the behavior of the curves in Fig.~3, which
correspond to a finite system, has a similar tendency to the CERES
data \cite{1,2}.


\begin{center}
\includegraphics[width=0.49\textwidth]{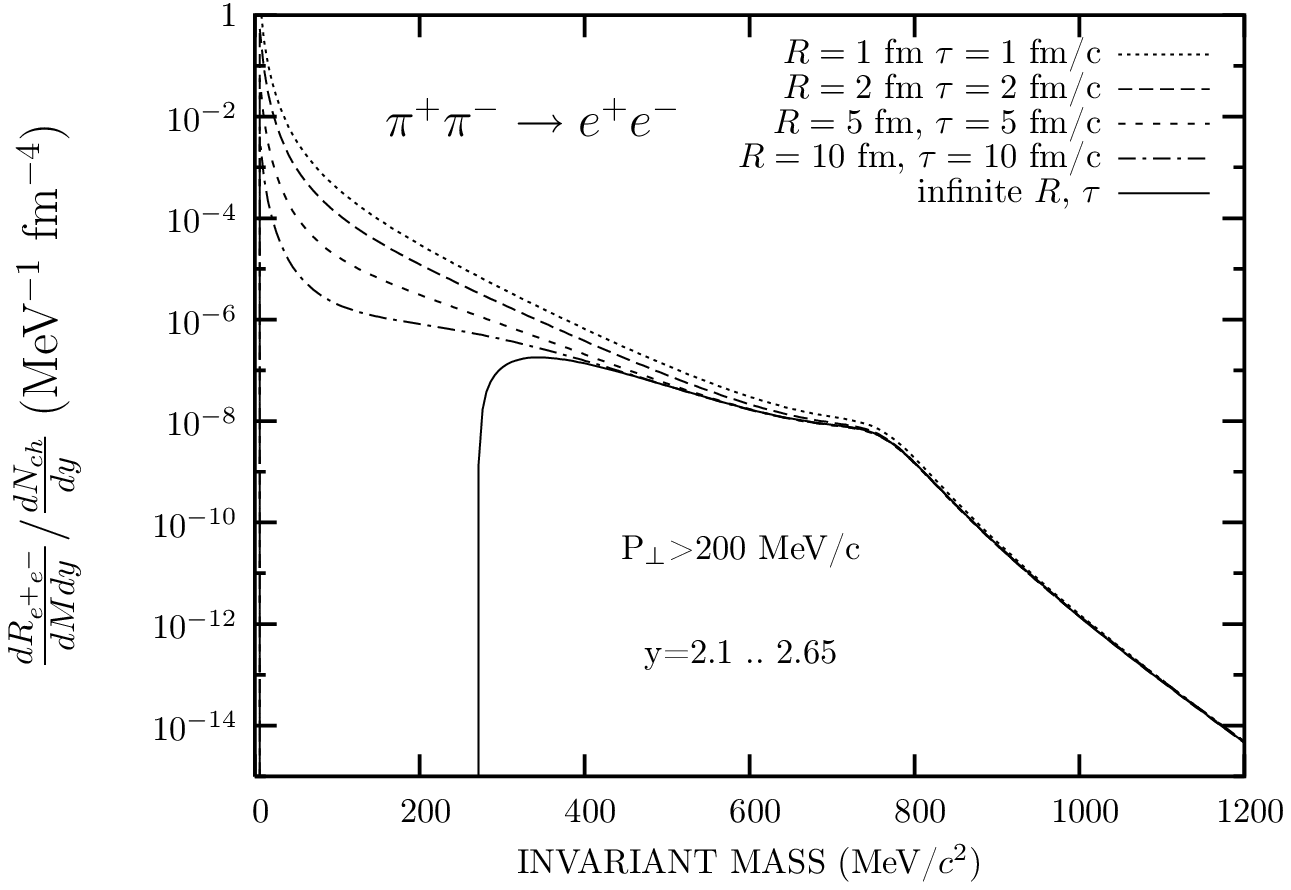}
\includegraphics[width=0.49\textwidth]{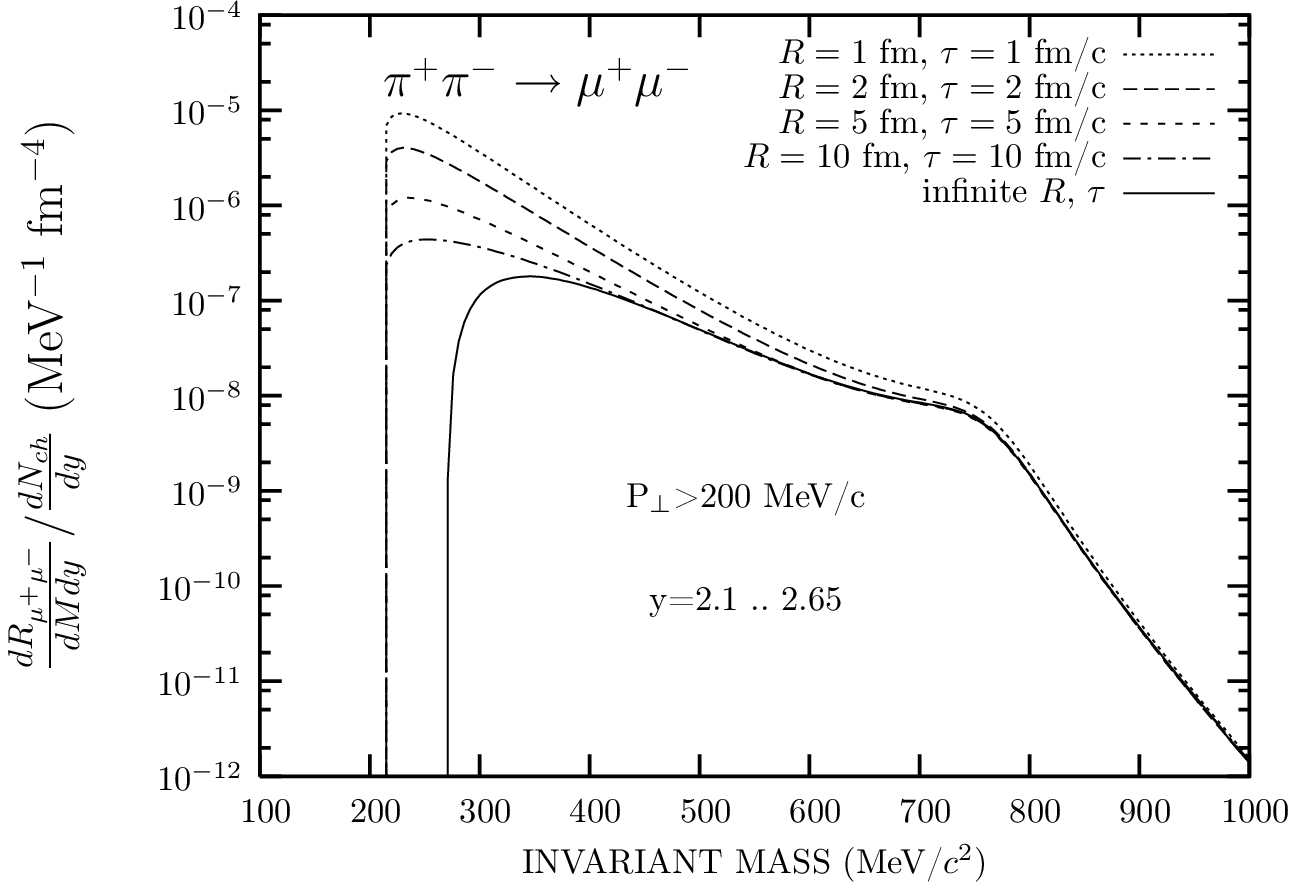}
\end{center}
\noindent {\footnotesize Fig. 3. Rates of electron-positron
(left panel) and muon-muon (right panel) productions in pion-pion
annihilation in a small finite system, $T=180$~MeV.}
\vspace*{10pt}%

For comparison, we present in Fig.~4 the results of evaluation of
the rate $dR^{(\rho)}_{e^+e^-}/dMdy$ of electron-positron pair
production in quark-antiquark annihilation in a QGP drop.
The evaluations was made under the same assumptions as for pion-pion
annihilation.
As in the previous case, one can see an increase in the rate with a
decrease in the invariant mass up to two electron masses.
This real threshold is close to the total mass of annihilating quarks
$M=2m_q\thickapprox 10$~MeV/c$^2$.
The different behavior of the rate for small parameters $R$ and $\tau$
in the region of small invariant masses $M\leqslant 500$~MeV/c$^2$, as
compared to the rate for infinite parameters $R=\infty$, $\tau=\infty$,
is due to an increase of quantum fluctuations which are evidently bigger
for a smaller size of the QGP drop.

One more important result is worth to note: the enhancement
of the dilepton production rate for the low invariant mass region is much
more sensitive to the variation in the spatial size of a many-particle
(pion, quark) system than to that in the system lifetime
(see Fig.~4, right panel). This
\begin{center}
\includegraphics[width=0.49\textwidth]{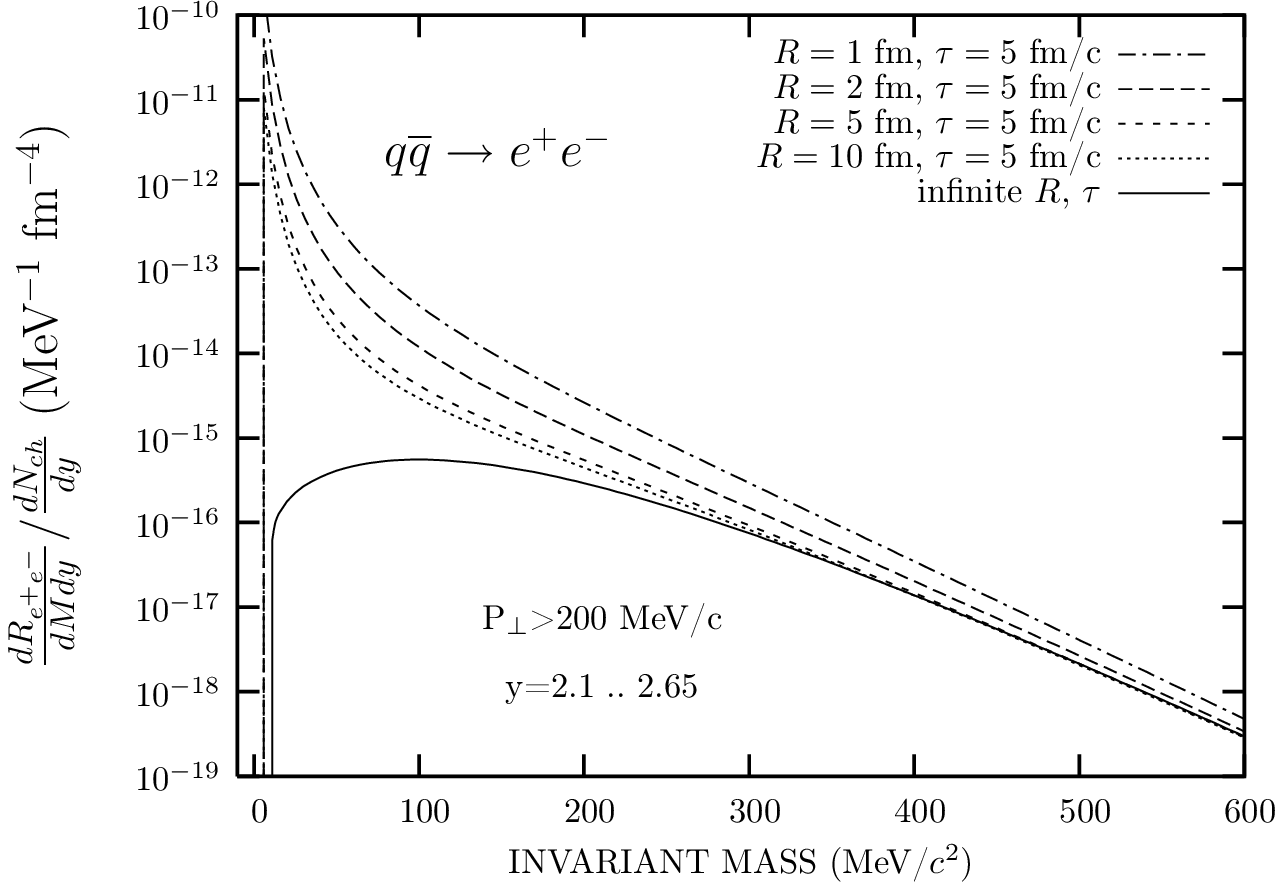}
\includegraphics[width=0.49\textwidth]{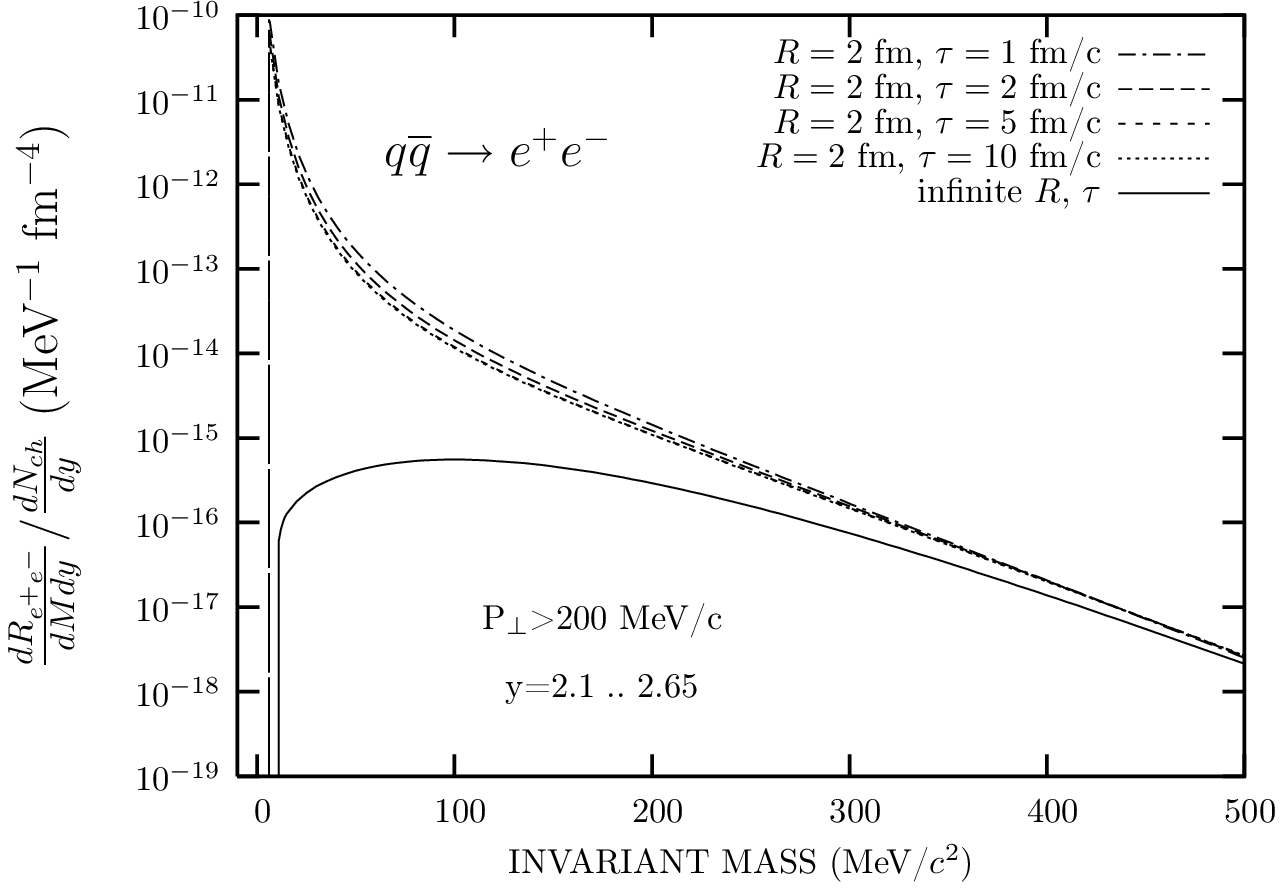}
\end{center}
\noindent {\footnotesize Fig. 4. Electron-positron
production rates in quark-antiquark annihilation in a hot QGP drop,
$T=180$ MeV.}
\vspace*{10pt}%

\noindent
fact can be explained with the use of the particle density taken in
a particular form (\ref{6}).
Indeed, for the system which is confined in three spatial dimensions,
each of the three spatial densities in (\ref{6}), for instance
$\exp{(-x_i^2/R^2)}$  with $i=1,2,3$, corresponds to one
integration in (\ref{1}) over a component of the pion-pair total
momentum $K_i$.
Hence, each of the three integrations on the r.h.s. of (\ref{2})
involves a factor which is responsible for quantum
fluctuations in its own spatial dimension; for instance, one of the three
factors is $(2\pi)^{1/2}R\cdot \exp{(-R^2K_i^2)}$.
Meanwhile, the spectral function which is
responsible for the system lifetime finiteness, for
instance $(2\pi)^{1/2}\tau\cdot \exp{(-\tau^2K_0^2)}$ as in our
approach, emerges as an integrand factor just once during the
integration over $K_0$.
That is why the 3-dimensional variations of the spatial size of a
many-particle system, as seen in Fig.~4,
influence the shape of the dilepton production rate much stronger
than the variations of the lifetime of a small system.

\medskip

{\bf Acknowledgements:}
We would like to thank D.~Blaschke and V.~Khryapa for
fruitful discussions.
D.A. wishes to express his deep appreciation to V.~Ruuskanen and
U.~Heinz for instructive discussions and support.


\end{document}